%% file: main.tex
\documentclass{article}
\usepackage{spconf,amsmath,graphicx}
\usepackage{multirow}
\usepackage{booktabs}
\usepackage{cite}
\usepackage{xcolor}
\usepackage{verbatim}
\usepackage{amsfonts}

\title{CTCBERT: Advancing Hidden-unit BERT with CTC Objectives}
\name{Ruchao Fan$^1$\sthanks{Work done during an internship at Microsoft.}, Yiming Wang$^2$, Yashesh Gaur$^2$, Jinyu Li$^2$}
\address{$^1$University of California, Los Angeles \quad $^2$Microsoft Corporation}
\begin{document}
\ninept
\maketitle
\begin{abstract}
In this work, we present a simple but effective method, CTCBERT, for advancing hidden-unit BERT (HuBERT). HuBERT applies a frame-level cross-entropy (CE) loss, which is similar to most acoustic model training. However, CTCBERT performs the model training with the Connectionist Temporal Classification (CTC) objective after removing duplicated IDs in each masked region. The idea stems from the observation that there can be significant errors in alignments when using clustered or aligned IDs. CTC learns alignments implicitly, indicating that learning with CTC can be more flexible when misalignment exists. We examine CTCBERT on IDs from HuBERT Iter1, HuBERT Iter2, and PBERT. The CTC training brings consistent improvements compared to the CE training. Furthermore, when loading blank-related parameters during finetuning, slight improvements are observed. Evaluated on the Librispeech 960-100h setting, the relative WER improvements of CTCBERT are 2\%-11\% over HuBERT and PERT on test-other data.
\end{abstract}

\input{Tex/intro}

\input{Tex/method}
\input{Tex/exp_setup}

\input{Tex/results}
\input{Tex/conclusion}

\vfill\pagebreak

\bibliographystyle{IEEEbib}
\bibliography{strings,refs}

\end{document}

%% file: Tex/intro.tex
\section{Introduction}
\label{sec:intro}
Recently, self-supervised learning (SSL) has gained lots of popularity because of its superiority of learning from large amounts of un-annotated data \cite{ao22_interspeech,ren22_interspeech,arunkumar22_interspeech,WangWCLLQY22,lee2022self,mohamed2022self,sriram22_interspeech}. This learning process is considered as \emph{pretraining}. Later on, fine-tuning a system using the pretrained model as a starting point often obtains impressive results\cite{zhang2022bigssl,chen2022wavlm,fan2022towards}. Such a good property motivates researchers to design SSL algorithms. The key of SSL is to construct a valuable task with un-annotated data in pretraining. Current SSL methods in speech can be categorized into either generative or discriminative approaches according to their loss functions. 

Generative approaches predict future frames given history context similar to GPT \cite{radford2018improving}, or reconstruct masked features given unmasked ones similar to BERT \cite{kenton2019bert}. Representative models in generative approaches include APC, VQ-APC \cite{ChungHTG19,ChungG20, Chung0G20}, DeCoAR \cite{LingLSK20}, Mockingjay \cite{LiuYCHL20}, TERA \cite{LiuLL21}, and MPC \cite{JiangLZCLHZHL21}. These approaches, however, perform predictions on a continuous space which may not be compatible with the fine-tuning tasks. On the other hand, discriminative approaches employ discriminative loss functions such that the model can be aware of the classification concepts, which may be beneficial for fine-tuning. Representative models in this category are Wav2vec \cite{SchneiderBCA19}, Wav2vec2.0 \cite{BaevskiZMA20}, HuBERT \cite{HsuBTLSM21} and PBERT \cite{wang2022supervision}. Wav2vec series discriminate negative samples from positive ones using a contrastive loss to push the prediction close to the positive sample and away from the negative samples. However, sampling negative samples is not always a well-defined problem because the definition of ``negative'' is sometimes vague and can vary depending on specific tasks. HuBERT \cite{HsuBTLSM21} uses an acoustic unit discovery system (K-means) to create pseudo-labels for each frame to avoid the process of sampling negative samples. However, HuBERT training needs multiple iterations to obtain a good performance. Instead of using k-means, PBERT \cite{wang2022supervision} uses a model trained with the finetuning data to obtain the IDs by conventional automatic speech recognition (ASR) decoding. The quality of these IDs are much better than that of clustered IDs in HuBERT. Hence, one iteration for PBERT can already obtain better performance than HuBERT.

In this paper, we propose a simple but effective method, CTCBERT, for training ID-based SSL systems including HuBERT and PBERT. Conventionally, HuBERT and PBERT use cross entropy (CE) loss in the training, which is similar to most acoustic model training. However, CTCBERT performs the model training with the Connectionist Temporal Classification (CTC) objective \cite{GravesFGS06} after removing duplicated IDs in each masked region. The idea is based on our observation that there are significant errors of alignments in learning IDs. CTC computes the summation of all possible alignments to labels and can learn the alignment implicitly. Hence, learning with the CTC objective can be more flexible when the misalignment exists. We examine CTCBERT on IDs from HuBERT Iter1, HuBERT Iter2, and PBERT. The CTC training has consistent improvements compared to the CE training, indicating that CTCBERT achieves better performance than HuBERT and PBERT, respectively. Another advantage of CTC training is that CTCBERT contains blank-related parameters that can potentially be used for finetuining. Slight improvements are observed when loading blank-related parameters from the pretrained model during finetuning.

The remainder of this paper is organized as follows. Section \ref{sec:method} introduces the proposed CTCBERT method. Experimental setups are described in Section \ref{sec:exp_setup}. Results are shown and discussed in Section \ref{sec:results}. We conclude the paper in Section \ref{sec:conclusion}.

%% file: Tex/method.tex
\section{CTCBERT}
\label{sec:method}
In this section, we introduce the proposed CTCBERT, from our motivations to the technical details. 

\subsection{Misalignment in Learning IDs}
\label{ssec:misalign_example}

\begin{figure}[t]
\centering
\centerline{\includegraphics[width=0.49\textwidth,height=0.38\textwidth]{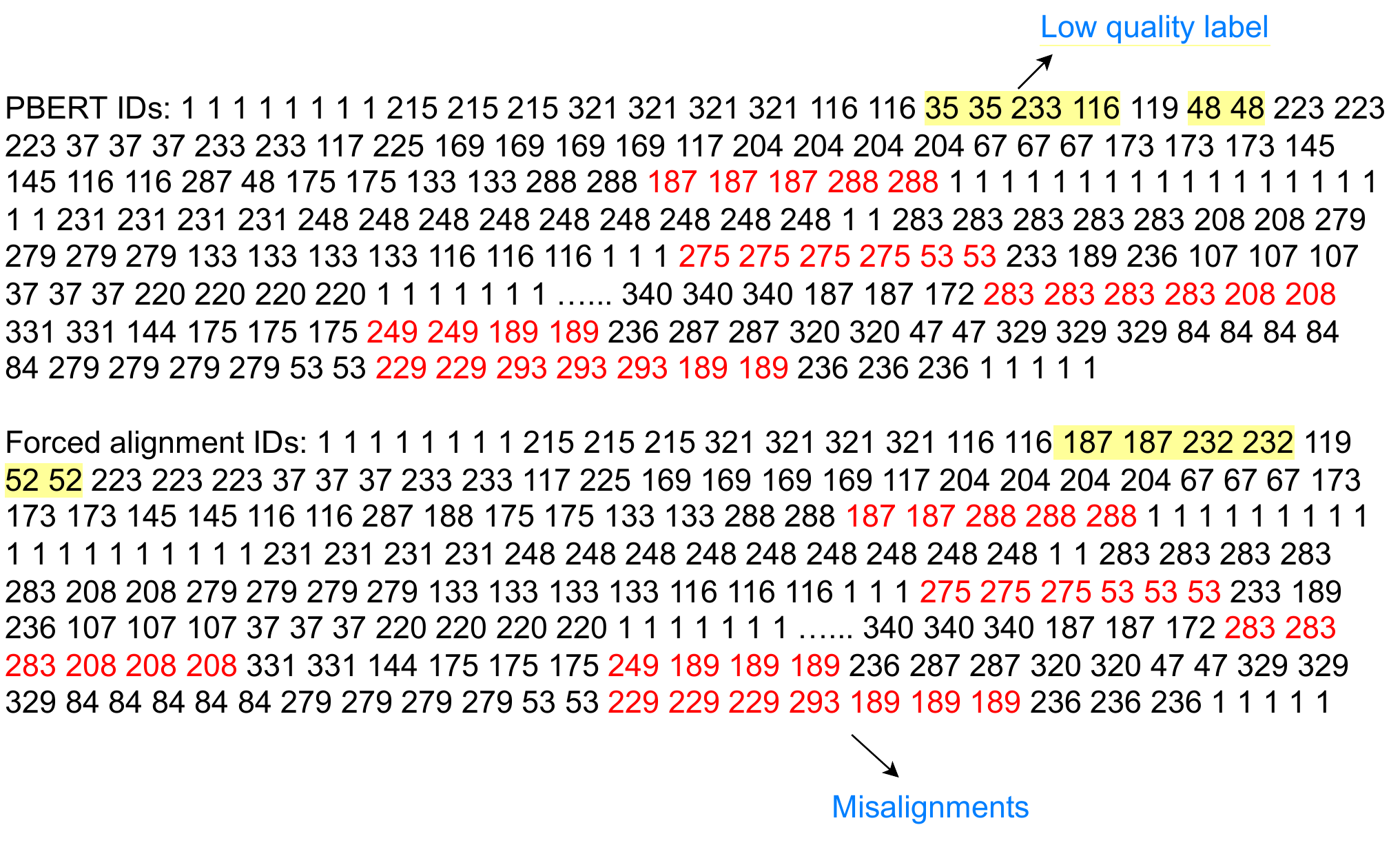}}
\caption{PBERT and forced alignment IDs. Forced aligned IDs are obtained from a hybrid model trained with train-960 data, and can be regarded as the ground-truth of the PBERT IDs. Yellow indicates the low quality labels. Red represents the misalignment.}
\label{fig:mismalign}
\end{figure}

The core of CTCBERT is to use the CTC objective for ID-based SSL pretraining. Hence, we first introduce our motivation to use the CTC objective. HuBERT and PBERT can achieve good performance for self-supervised learning. Are they reaching the top performance? How can we further improve the performance of self-supervised learning? The simplest solution may be increasing the quality of the learning IDs by more iterations. However, multiple pre-training iterations would incur more computational cost and is not environmentally friendly.

Recently, we analyze the obtained IDs from PBERT and compare them with the ground-truth IDs. The ground-truth IDs are acquired by training a hybrid model with the entire train-960 data first. Then we use the ground-truth to perform the forced-alignment, and obtain the forced-aligned IDs, which, therefore, can be regarded as the ground-truth PBERT IDs. The comparisons of the two types of IDs are shown in Fig.\ref{fig:mismalign}. As we can observe from the figure, there are two main errors in PBERT IDs. One is the low quality label in yellow. The other is the misalignment in red. The quality of label is hard to improve. However, it may be feasible to mitigate the misalignment problem. By carefully observing each misalignment in red in Figure \ref{fig:mismalign}, we hypothesize that CTC is a more appropriate loss function than CE loss for the misalignment problem. For example, the PBERT IDs are $\{187,187,187,288,288\}$ and the ground-truth IDs are $\{187,187,288,288,288\}$. If CTC loss is used, the targets are both $\{187,288\}$. The alignment can be learned implicitly in the model with CTC loss. The model will be more flexible and tolerant to the misalignment, which is our motivation to train the ID-based SSL systems with the CTC objective.

\subsection{HuBERT}
\label{ssec:hubert}
HuBERT leverages K-means to create pseudo-labels for each frame, and then uses the pseudo-labels to pretrain the model. HuBERT consists of a feature extractor, a transformer encoder, a final projection layer and clustered ID embeddings. Mathematically, suppose the input raw waveform is $X=\{x_1,x_2,...,x_n\}$, then after the feature extractor layers, we obtain the feature sequence $Z=\{z_1,z_2,...,z_T\}$. HuBERT masks $Z$ sequence to predict masked regions given unmasked regions. Let masked regions be $Z_{mask}$. $Z$ with masked embeddings is fed to the transformer encoder $H$, and then the projection layer $P$. After computing the similarity between the projection layer outputs and the clustered ID embeddings $\{E_0, E_1,...,E_{N-1}\}$, we finally obtain the softmax output for each frame, denoted as: $y=\{y_1,y_2,....,y_T\}$.

Suppose the pseudo-label IDs are $C=\{c_1,c_2, ..., c_T\}$, then HuBERT loss can be described as:

\begin{equation}
\label{eq:hubertloss}
\begin{aligned}
L_{\text{HuBERT}} = -\sum_{t\in Z_{mask}}logP(c_t|Z) \\
\end{aligned}
\end{equation}
HuBERT is essentially using cross-entropy (CE) loss to maximize the logits for each frame and improves the correctness of the predictions. As a result, the low quality labels would have a negative effect for the HuBERT pretraining.

\subsection{PBERT}
\label{ssec:pbert}
Instead of using the acoustic unit discovery system to create pseudo-labels for each utterance, PBERT uses a small amount of paired data (finetuning data) to train a hybrid model first, and then apply decoding on the training data to get the aligned IDs. The quality of the aligned IDs, from the perspective of speech recognition, is better than the clustered IDs. Hence, the performance of PERT is better than that of HuBERT, even with only one iteration. The pretraining model structure of PBERT is the same as HuBERT. PBERT uses CE loss as well to increase the prediction accuracy of each frame.

If using the CE loss in HuBERT and PBERT, the misalignment errors can be regarded as the low quality label. And CE loss will try to increase the accuracy with the low quality label, which hurts the model training. As mentioned in Section \ref{ssec:misalign_example}, using CTC loss may solve the problem of the misalignment.

\subsection{Connectionist Temporal Classification}
\label{ssec:ctc}
Connectionist Temporal Classification (CTC) is a well-known sequence-level loss for speech recognition. CTC adds the ``blank'' token to solve the length mismatch problem between the input sequence and the label sequence. Suppose the input sequence is $X'=\{x'_1, x'_2, .... ,x'_T\}$ and the label sequence is $Y=\{Y_1, Y_2, ...., Y_u\}$. Define a mapping rule $\beta$ that removes blank and repetitive tokens of the input sequence. Then, the CTC can be described as a summation of all the paths that can be mapped by rule $\beta$ to the label $Y$. We can write the CTC loss function as:

\begin{equation}
\label{eq:ctc}
\begin{aligned}
L_{\text{CTC}} = -\sum_{a\in \beta^{-1}(Y)} \prod_{t=1}^T P(a_t|X')
\end{aligned}
\end{equation}
where $\beta^{-1}$ is the inverse mapping function. CTC learns the alignment between the input and output sequence. It does not affected by the misalignment. We use another example in Figure \ref{fig:mismalign}: the aligned IDs $\{229, 229, 293, 293, 293, 189, 189\}$ and the ground-truth IDs $\{229, 229, 229, 293, 189, 189, 189\}$ are basically two paths of the label sequence $\{229, 293, 189\}$ and are both considered in the CTC loss. Hence, CTC loss is tolerant and not sensitive to the misalignment. Consequently, we assume that using CTC loss can well solve the problem of misalignment.

\subsection{CTCBERT}
\label{ssec:ctcbert}
Using the example in the section \ref{ssec:ctc}, we can easily find that if CTC loss is used, the targets are mapped to $\{229, 293, 189\}$ for both ID sequences. CTC can take care of the alignment in the sequence during training, indicating that training with CTC is more flexible and tolerant to the misalignment. Compared to CE training, which is frame-based and is very sensitive to label assumptions for each frame, CTC training is utterance/segment-based and is not sensitive to the frame alignment. Suppose the mask regions are $Z_{mask}=\{Z_1, Z_2, ..., Z_M\}$ and the corresponding masked IDs are $C=\{C_1, C_2, ...., C_M\}$. We define a rule $A$, which is to remove the repetitive IDs in each masked region. The loss function of CTCBERT can be described as:

\begin{equation}
\label{eq:ctcbert}
\begin{aligned}
L_{\text{CTCBERT}} =
- \sum_{m=1}^M \sum_{a\in \beta^{-1}(A(C_m))} \prod_{t\in Z_m} P(a_t|Z) \\
\end{aligned}
\end{equation}
where $\beta^{-1}$ is the inverse function of $\beta$ that maps all the paths to $A(C_m)$. However, in our preliminary experiments, we find that CTC training only improves on HuBERT Iter2 IDs. In other cases, we observe slow convergence of CTC training. Hence, we borrow canonical ideas from acoustic model training \cite{ZhangL18}: CTC and CE joint training or using CE warmup. For CTC and CE joint training, we set the task ratio of CTC loss as $\alpha$, the loss function can be rewritten as:
\begin{equation}
\label{eq:joint training}
\begin{aligned}
L_{\text{joint}} = \alpha L_{\text{CTCBERT}} + (1-\alpha) L_{\text{HuBERT}}
\end{aligned}
\end{equation}
For CE warmup, it represents that the model is trained with CE loss for certain steps at the beginning, and then the model aims for CTC training. Both solutions can alleviate the slow convergence phenomena of CTC training. Another advantage of training ID-based SSL systems with the CTC loss may be the consistency of the pretraining and finetuning task (when using the CTC objective).

\subsection{Blank-Related Parameters}
\label{ssec:blank_param}
Computing the CTC loss involves an additional blank token, which is the same as the finetuning task. Therefore, we are thinking whether the blank-related parameters used in the pretraining can be used for the finetuning task. When we use the CTC loss, the ID embeddings are $\{E_b, E_0, E_1,...., E_{N-1}\}$, where $E_b$ is the embebdding for the blank token. If the weight of the projection layer is $W_p \in \mathbb{R} ^ {d_{embed} \times d_{model}} $ and its bias is $b_p \in \mathbb{R} ^ {d_{embed}}$, we can extract the blank-related parameters as follows:

\begin{equation}
\label{eq:blank_param}
\begin{aligned}
W_b &= W_p ^ T * E_b \in \mathbb{R}^{d_{model}} \\
b_b &= b_p * E_{blank} ^ T \in \mathbb{R}
\end{aligned}
\end{equation}
$W_b$ and $b_b$ are then used for initializing the corresponding blank-related parameters in the last projection layer during finetuning. Note that this is effective when CTC loss is used during finetuning. 

%% file: Tex/exp_setup.tex
\section{Experimental Setup}
\label{sec:exp_setup}

\subsection{Dataset}
\label{ssec:dataset}
In this paper, we evaluate our method on the LibriSpeech 960h-100h benchmark \cite{PanayotovCPK15}. Librispeech contains 960 hours of training data from read audiobooks and is being split into three subsets: train-clean-100, train-clean-360 and
train-other-500. We use the entire 960h data as the pretraining data (without transcripts), and use the train-clean-100 subset as the finetuning data (with transcripts).

\subsection{Pretraining}
\label{ssec:pretraining}
We conduct experiments on IDs from HuBERT Iter1, HuBERT Iter2, and PBERT with the same model architecture (HuBERT-Base) in \cite{HsuBTLSM21}. For HuBERT Iter1 and HuBERT Iter2, their clustered IDs are obtained from K-means of MFCC and HuBERT intermediate layer features, respectively. For PBERT, the train-clean-100 portion is used to train a hybrid model. The acoustic model is a TDNN-F \cite{PoveyCWLXYK18} trained with LF-MMI criterion \cite{PoveyPGGMNWK16}. We decode the entire train-960 data using the 3-gram LM in the Librispeech data, and then rescore the lattices using the 4-gram LM, which is also in the Librispeech data. The best path is converted to the ID sequence as the PBERT IDs. We use 347 distinct positional-dependent phonemes as the training targets. All models are trained on 32 GPUs, with a batch size of at most 87.5 seconds of audio per GPU for 400k steps. The mask strategies are the same for all model training including HuBERT, PBERT and CTCBERT. Specifically, the mask span is set to $l=
10$, and $p = 8\%$ of the waveform encoder output frames are randomly selected as mask start. We use AdamW optimizer \cite{loshchilov2018decoupled} with weight decay 0.01 and $\beta = (0.9, 0.98)$. The learning rate ramps up linearly for the first 32k steps and then decays linearly back to 0. The peak learning rates is 5e-4. We replicate the results in the literature.

\subsection{Finetuning}
\label{ssec:finetuning}
The model is fine-tuned on 8 GPUs with a batch
size of 200 seconds of audio per GPU for 80k steps. The convolutional feature extractor is always fixed. Adam optimizer is used for all model training. A tri-stage scheduler of the learning rate is used, where the first 10\% of updates are for warm up, the learning rate held constant for the next 40\% and then linearly decayed for the rest of updates. The evaluation is based on the wav2letter++ \cite{pratap2019wav2letter++} beam search decoder wrapped in Fairseq \cite{ott2019fairseq}. For language model-fused decoding, a beam size 1500 and a 4-gram LM are used.

%% file: Tex/results.tex
\section{Results and Discussion}
\label{sec:results}

\subsection{Main Results}
\label{ssec:main_results}

\input{Tables/main_results}
In this section, we present the performance of CTCBERT on three different IDs: HuBERT iter1, HuBERT iter2, and PBERT. The results are shown in Table \ref{tab:main_results}. For HuBERT iter1 IDs, CTCBERT improves the WER from 16.5\% to 15.8\% on the test other data when no language model is used. The improvement becomes smaller ($9.5\%\rightarrow 9.4\%$) when the 4-gram language model is used during decoding. Using HuBERT iter2 IDs, CTCBERT achieves the best improvements. CTCBERT obtains 11\%, and 9\% relative WER improvements compared to CE training when no language model and 4-gram language model is used, respectively. For PBERT, CTCBERT also achieves improvements from 7.7\% to 7.5\% on test-other data. Overall, the improvements on noisy data are larger than that on clean data. The reason may be because noisy speech are more likely to generate IDs with misalignments. CTCBERT achieves better performance than HuBERT or PBERT. Note that we replicated the results in the literature, and may cause difference in WER. However, the comparisons are reasonable because the same target IDs are used for CE and CTC training.

We can observe from Table\ref{tab:main_results} that when using blank-related parameters during finetuning, most of the models get further improvements. For example, for HuBERT iter1, the WER drops from 9.4\% to 9.3\%, and for PBERT, the WER drops from 7.5\% to 7.4\%, all on the test other data. We do not see an improvement for HuBERT iter2. The reason may be that the performance of CTCBERT has already been very good. However, we do not see loading blank-related parameters in finetuning hurt the performance a lot. Loading blank-related parameters overall brings positive effect.

\subsection{Stabilize the CTC training}
\label{ssec:ce_ctc_joint}
As we mentioned in Section \ref{ssec:ctcbert}, CTC sometimes shows slow convergence, and needs strategies like CTC and CE joint training, or CE warmup to stabilize the training. For different IDs: HuBERT iter1, HuBERT iter2, PBERT, their best performance achieves differently. The results of the effect of CE warmup and CTC, CE joint training are shown in Table \ref{tab:cectcjoint}. We can observe from the table that for CTCBERT with HuBERT iter1 IDs, CE warmup training can improve the performance. However, the best performance achieves at $0.5\text{CE}+0.5\text{CTC}$. For CTCBERT with HuBERT iter2 IDs, the pure CTC training achieves the best performance. For CTCBERT with PERBT IDs, the best performance also achieves at $0.5 \text{CE} + 0.5 \text{CTC}$. In fact, $0.5 \text{CE} + 0.5 \text{CTC}$ also achieves improvements on HuBERT iter2 IDs compared to CE training. 

Hence, how to choose the training strategie and the ratio in CE and CTC joint training? From our experience, CTC and CE joint training is always better than CE warmup training. For CTC and CE joint training, starting CTCBERT with the ratio of $0.5 \text{CE} + 0.5 \text{CTC}$ would be a good choice.

\input{Tables/stabletraining}

\subsection{Quantitative Analysis}
\label{ssec:qual_analysis}
\begin{figure}[t]
\centering
\centerline{\includegraphics[width=0.45\textwidth,height=0.33\textwidth]{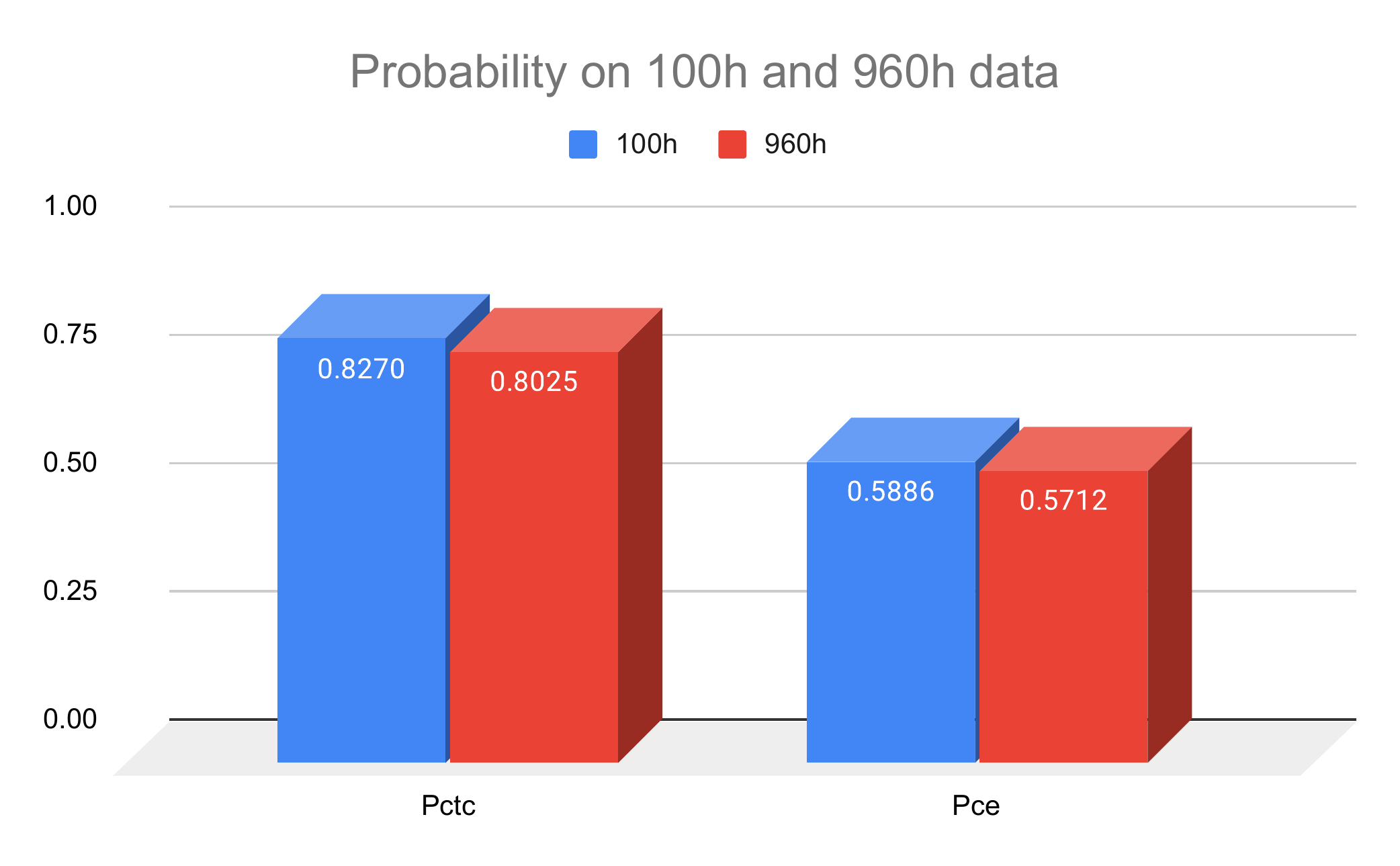}}
\caption{Posterior probabilities of CTC and CE models calculated on train-clean-100h and train-960h data. IDs are obtained from force alignment using ground-truth and a hybrid model trained with train-clean-100 data. Hence, the IDs quality of train-clean-100h data is better than that of train-960 data.}
\label{fig:quan_analysis}
\end{figure}

We want to perform a quantitative analysis to again confirm that CTC is more tolerate to misalignment of clustered and force aligned IDs. To do so, we first train a hybrid model with the train-clean-100h data. Then, using ground-truth and the hybrid model, we obtain the forced aligned IDs for train-960h and train-clean-100h data. Hence, the quality of alignment on train-clean-100h data is better than that of 
train-960h data. If CTC is more tolerate to misalignment, the model trained with the CTC loss should provide a smaller degradation of the posterior probability for data (an average probability of all ground-truth tokens) from train-clean-100h to train-960h than the model trained with CE loss. The results are shown in Fig.\ref{fig:quan_analysis}. When we compute the relative posterior probability degradation, we get $(0.8270-0.8025)/0.8270 = 2.9\%$ for the CTC model, and $ (0.5886-0.5712)/0.5886 = 3.1\%$ for the CE model. The CTC model has a smaller degradation of posterior probability from train-clean-100 to train-960 than the CE model. It confirms our conclusion that CTC is more tolerant to the misalignment. Note that the relative improvements of probability from CE to CTC ($3.1\% \rightarrow 2.9\%$) are similar to the relative WER improvements ($7.7\% \rightarrow 7.4\%$) when training PBERT with the CTC objective.

%% file: Tables/main_results.tex
\begin{table}[tp]
\caption{The performance of CTCBERT using HuBERT iter1, HuBERT iter2, and PBERT IDs. "load blk." indicates loading blank-related parameters during finetuning.}
\scriptsize
\vspace*{5mm}
\centering
\begin{tabular}{c c cccc}
\hline
\multirow{2}{*}{Model} & \multirow{3}{*}{LM} & \multicolumn{2}{c}{dev} & \multicolumn{2}{c}{test} \\
\cmidrule(r){3-4} \cmidrule(r){5-6} 
~ & ~ & clean &  other &  clean &  other\\
\hline \hline
\multirow{2}{*}{Supervised baseline} & None & - & - & 8.8 & 26.5  \\
~ & 4-gram  & - & - & 5.0 & 16.8  \\
\hline
\multirow{2}{*}{HuBERT iter1\cite{HsuBTLSM21}} & None & 7.1 & 16.7 & 7.2 & 16.5  \\
~ & 4-gram  & 3.2 & 9.6 & 3.8 & 9.5  \\
\hline
\multirow{2}{*}{$\rightarrow$ CTCBERT} & None & 6.9 & 15.9 & 7.0 & 15.8  \\
~ & 4-gram & 3.2 & 9.2 & 3.8 & 9.4  \\
\multirow{2}{*}{ + load blk.} & None & 6.8 & 15.8 & 6.9 & 15.7  \\
~ & 4-gram & 3.2 & 9.2 & 3.7 & 9.3  \\
\hline
\multirow{2}{*}{HuBERT iter2\cite{HsuBTLSM21}} & None & 5.3 & 13.7 & 5.5 & 13.6  \\
~ & 4-gram  & 2.8 & 8.7 & 3.4 & 8.7  \\
\hline
\multirow{2}{*}{$\rightarrow$ CTCBERT} & None & 5.2 & 12.3 & 5.4 & 12.1  \\
~ & 4-gram & 2.7 & 7.8 & 3.3 & 7.9  \\
\multirow{2}{*}{ + load blk.} & None & 5.2 & 12.4 & 5.4 & 12.1  \\
~ & 4-gram & 2.7 & 7.9 & 3.3 & 7.9  \\
\hline
\multirow{2}{*}{PBERT\cite{wang2022supervision}} & None & 4.6 & 11.7 & 4.8 & 11.8  \\
~ & 4-gram  & 2.6 & 7.3 & 3.2 & 7.7  \\
\hline
\multirow{2}{*}{$\rightarrow$ CTCBERT} & None & 4.7 & 11.5 & 4.8 & 11.4  \\
~ & 4-gram & 2.5 & 7.1 & 3.2 & 7.5  \\
\multirow{2}{*}{ + load blk.} & None & 4.6 & 11.3 & 4.8 & 11.3  \\
~ & 4-gram & 2.5 & 7.1 & 3.1 & 7.4  \\
\hline
\end{tabular}
\label{tab:main_results}
\end{table}


%% file: Tables/stabletraining.tex
\begin{table}[tp]
\caption{\small {The effect of CE warmup and CTC, CE joint training.}}
\scriptsize
\vspace*{5mm}
\centering
\begin{tabular}{l cc cccc}
\hline
\multirow{2}{*}{Model} & \multicolumn{2}{c}{ce} &  \multicolumn{2}{c}{dev} & \multicolumn{2}{c}{test} \\
\cmidrule(r){2-3} \cmidrule(r){4-5} \cmidrule(r){6-7} 
~ & ratio & warmup &  clean &  other &  clean &  other\\
\hline \hline
\multirow{5}{*}{\shortstack{CTCBERT \\with iter1 IDs}} & 0.0 & 0 & 7.2 & 16.8 & 7.5 & 16.7  \\
~ & 0.0 & 32k & 7.2 & 15.8 & 7.3 & 15.8  \\
~ & 0.2 & 0  & 7.2 & 16.6 & 7.3 & 16.7  \\
~ & 0.5 & 0  & 6.9 & 15.9 & 7.0 & 15.8  \\
~ & 0.8 & 0  & 6.8 & 16.0 & 7.0 & 16.1  \\
~ & 1.0 & 0  & 7.1 & 16.7 & 7.2 & 16.5  \\
\hline
\multirow{3}{*}{\shortstack{CTCBERT \\with iter2 IDs}} & 0.0 & 0 & 5.2 & 12.3 & 5.4 & 12.1  \\
~ & 0.0 & 32k & 5.2 & 12.6 & 5.4 & 12.4  \\
~ & 0.5 & 0  & 5.5 & 13.3 & 5.6 & 13.1  \\
~ & 1.0 & 0  & 5.3 & 13.7 & 5.5 & 13.6  \\
\hline
\multirow{3}{*}{\shortstack{CTCBERT \\with PBERT IDs}} & 0.0 & 0 & 5.5 & 13.8 & 5.7 & 13.9  \\
~ & 0.0 & 150k & 4.7 & 11.6 & 4.9 & 11.7  \\
~ & 0.5 & 0 & 4.7 & 11.5 & 4.8 & 11.4  \\
~ & 1.0 & 0 & 4.6 & 11.7 & 4.8 & 11.8  \\
\hline
\end{tabular}
\label{tab:cectcjoint}
\end{table}


%% file: Tex/conclusion.tex
\section{Conclusion}
\label{sec:conclusion}

In this paper, we present a simple but effective method,
CTCBERT, for advancing hidden-unit BERT (HuBERT). HuBERT uses a frame-level cross-entropy loss in the masked regions, which is similar to most acoustic model training. In contrast, the proposed CTCBERT performs model training with CTC objectives after removing duplicated IDs in each
masked region. The idea stems from our observation that
there can be significant errors in alignments when using
clustered or aligned IDs. Through our quantitative analysis, we confirm that CTC is more tolerant to misalignment of IDs. This implies learning with the CTC objective can be more flexible when misalignment exists. CTCBERT is examined on the three different IDs: HuBERT Iter1, HuBERT Iter2, and PBERT. From our experiments, CTCBERT achieves consistent improvements on the three IDs compared to the CE training. Furthermore, slight improvements are observed, when loading blank related parameters during the finetuning. Under the setting of Librispeech 960h-100h, CTCBERT on HuBERT iter 2 achieves 11\% relative WER improvements compared to CE training. For other IDs, CTCBERT can have a consistent relative WER improvements of around 4\% compared to CE training. Empirically, experimenting CTCBERT with $0.5 \text{CE} + 0.5 \text{CTC}$ would be a good starting point for stabilizing the training.